# On the Comparison of Single-Carrier vs. Digital Multi-Carrier Signaling for Long-Haul Transmission of Probabilistically Shaped Constellation Formats

Kaoutar Benyahya[1], Amirhossein Ghazisaeidi[1], Vahid Aref[2], Mathieu Chagnon[2], Aymeric Arnould[1], Stenio Ranzini[2], Haik Mardoyan[1], Fred Buchali[2] and Jeremie Renaudier[1]

*(1) Nokia Bell Labs Paris-Saclay, Route de Villejust, Nozay, 91620 France (Kaoutar.benyahya@nokia-bell-labs.com)*
*(2) Nokia Bell Labs Stuttgart, Lorenzstraße 10, 70435 Stuttgart Germany*

**Abstract:** We report on theoretical and experimental investigations of the nonlinear tolerance of single carrier and digital multicarrier approaches with probabilistically shaped constellations. Experimental transmission of PCS16QAM is assessed at 120 GBd over an ultra-long-haul distance.

## 1. Introduction

Coherent optical communication systems have helped to keep up with the exponential growth of bandwidth-hungry services over the internet for over a decade. Digital signal processing (DSP) was the enabler of the first coherent 100 Gb/s solutions in the early 2010s, and there has been continued research efforts in advanced DSP techniques to enable higher bit rates and transmission reach, with the help of higher order modulation formats, efficient linear and nonlinear impairment mitigation schemes, and powerful forward error correction (FEC) algorithms. The advent of probabilistic constellation shaping (PCS) in 2015 [1-2] has paved the way for a higher degree of flexibility in transponders, now able to optimize the signal spectral efficiency with quasi infinite granularity as a function of the amount of noise and reach. Continuous technological improvements in transceiver design have also allowed to push the symbol rates of coherent transceivers north of 100 GBd [3-4], further improving the overall spectral efficiency as less guard-bands spectral gaps are needed to fully populate the C-band with carriers, themselves carrying signals of higher spectral efficiencies. Moreover, a lesser number of carriers simplifies the network deployment and management and reduces the number of transponders. Digital multicarrier modulation has attracted much attention lately for high symbol rate transponders [3], deemed as a potential scheme to reduce the transmission impairments such as Kerr nonlinearity [5] and electronically enhanced phase noise (EEPN) [6,7] for signals transmitted over ultra-long-haul distances. In this paper, we theoretically and experimentally investigate the transmission impairments of both single carrier and digital multi-carrier systems employing probabilistically shaped constellations and operating at a high aggregate symbol rate of 120 GBd. We compare numerical predictions of the nonlinear variance of both modulation techniques carrying PCS16QAM and PCS64QAM over ultra-long-haul distances. We experimentally assess the performance of these modulation methods when carrying PCS16QAM signals over submarine distances up to more than 10000 km.

## 2. Numerical investigation

Poggiolini et al. calculated in [5] that multiplexing several digital subcarriers of lower symbol rates can provide reach improvements when using a quadrature phase shift keying (QPSK) format, although this gain appeared dependent on the WDM bandwidth considered [8]. Nowadays flexible transceivers often use capacity approaching solutions based on probabilistic constellation shaping, as illustrated in Fig. 1-a). Contrary to QPSK or regular QAM constellations, PCS shapes the probability of occurrences of constellation points to follow a Gaussian profile which may induce different behavior in term of nonlinear tolerance. Although a first experimental investigation has been reported in single channel transmission [9], the joint impact of PCS applied to digital multi-carrier modulation remains unclear in WDM regime. We carried out numerical investigations to fully understand the nonlinear tolerance of multicarrier (MC) and single carrier (SC) signals modulated by PCS formats. We performed full C-band propagation simulations of 120GBd signals in 150GHz slots using split-step Fourier method (SSFM) using fiber parameters matching the actual fiber employed in the subsequent experimental demonstration. We computed the nonlinear coefficients $a_{nli}$ of the received constellations as a function of distance for SC (M=1) and MC (M=8), where M is the number of digital subcarriers, having the same aggregate symbol rate: $1 \times 120$ GBd or $8 \times 15$ GBd. The nonlinear coefficient $a_{nli}$ characterizes the variance of the nonlinear distortions, considered to be an additive Gaussian noise source, as $\sigma_N^2 = a_{nli}P^3$, where $P$ is the channel averaged launched power. We assume $P$ to be the optimal power corresponding to the nonlinear threshold (NLT) and assume the same PCS entropy for all subcarriers. ASE and laser linewidth were set to zero. Fig.1 b) and c) illustrate that for both PCS16QAM (entropy of 3) and PCS64QAM (entropy of 5), the nonlinear coefficients are very close for SC and MC no matter the distance, indicating similar nonlinear tolerance for both schemes. This result is attributed to the kurtosis of the PCS formats on the nonlinear variance as expected from the perturbation theory.

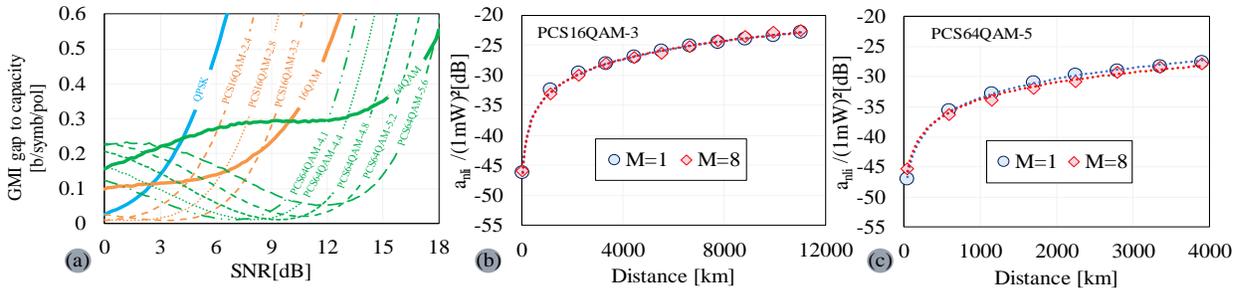

Fig. 1: (a) GMI gap to capacity versus SNR (b) Example of numerical nonlinear coefficient $a_{nli}$ versus distance for PCS16QAM and (c) PCS64QAM for M=1 and M=8, M: number of digital subcarriers.

## 3. Experimental setup and results

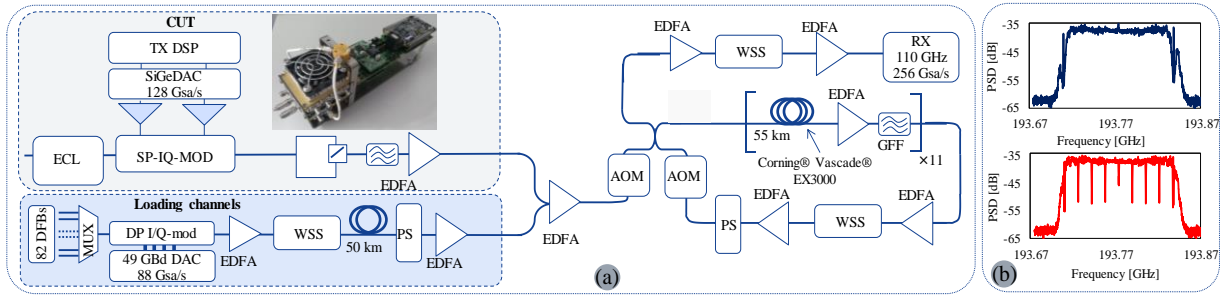

Fig. 2: (a) Experimental Setup (b) Optical spectrums of single-carrier and multi-carrier signals at the output of the transmitter of the CUT

We conducted an experimental comparison of the nonlinear tolerances of single-carrier and 8-digital subcarrier signals in a fully populated C-band scenario. The experimental setup is depicted in Fig. 2-a). The transmitter is made of 82 loading channels on a 50 GHz grid and a channel under test (CUT) occupying 3 slots of 50 GHz, being either the 120 GBd SC or the 8-MC signal. The CUT is generated using two 128 GSa/s SiGe DACs having ~4 bits of effective resolution and 24 GHz of 3-dB bandwidth [10]. After applying a linear digital pre-emphasis, clipping and quantization, the two (In-phase/Quadrature) tributaries are fed to the DACs. Amplifiers of 60 GHz 3-dB bandwidth are connected to the DAC's outputs, driving a Lithium-Niobate IQ modulator of about 41 GHz 3-dB bandwidth. The optical carrier is generated by an ITLA laser source at 193.775 THz. Polarization multiplexing is emulated using a delay-rotate-combine stage having a decorrelation delay of 27 ns. Optical pre-emphasis is performed with programmable optical filter. Fig. 2-b) shows the output spectra for SC and MC. In order to ease the comparison, all subcarriers of an MC modulated CUT have the same QAM modulation format and constellation entropy as used for the SC case, i.e. PCS16QAM of entropy 3 bits/symbol. The CUT is multiplexed with 82 loading channels, each operating at 49 GBd. The WDM signal is then amplified by a final EDFA before being sent into the recirculating loop. The transmission loop is made of 11 spans of 55 km EX3000 fiber, each followed by an EDFA. The launch power is set at 19 dBm, corresponding to the NLT of the system, maximizing the linear-nonlinear balance. At the output of the 11 spans, we use a 50 GHz-grid-resolution WSS for channel power equalization. A loop-synchronous polarization scrambler (PS) is used to randomly rotate the state of polarization after each recirculation in the loop. At the receiver (RX), the CUT is extracted using a WSS, amplified and sent to the coherent receiver. Electrical waveforms are acquired with a 110 GHz high speed real-time sampling scope operating at 256 GSa/s. Offline digital signal processing (DSP) is performed to process the received streams of 10 million samples. In case of MC, we demultiplex the different carriers before applying standard DSP per carrier. After chromatic dispersion compensation, polarization demultiplexing is performed, followed by carrier frequency and phase estimation using 2% pilot symbols. The lab transmitter setup suffers from I/Q timing skew effects, long echoes, and impedance mismatch due to long RF cable assemblies combined with non-identical RF matching. To compensate for that, a real-valued MIMO is added to the coherent DSP stack for both MC and SC modulation schemes before estimating the SNR. In the case of MC, an 8x8 real-valued MIMO equalization is employed, where the 8 inputs are the 4 inputs of each mirror sub-carrier in the MC spectrum, and the 8 outputs are equalized versions of those. On the other hand, a simpler 4x4 MIMO is applied for SC, as SC has no spectral 'mirror'. It is known that digital multi-carrier is more sensitive to transmitter impairments than single carrier [11]. As a consequence, and to be fair with both formats, the real-valued MIMO was adapted in data-aided mode for both MC and SC. Experimental results are presented in Fig.3. Fig.3-a) shows the measured average SNR with increasing distances for single carrier and multicarrier formats. The figure also shows the theoretical SNR at optimum power for both SC and MC signals computed according to the following equation:

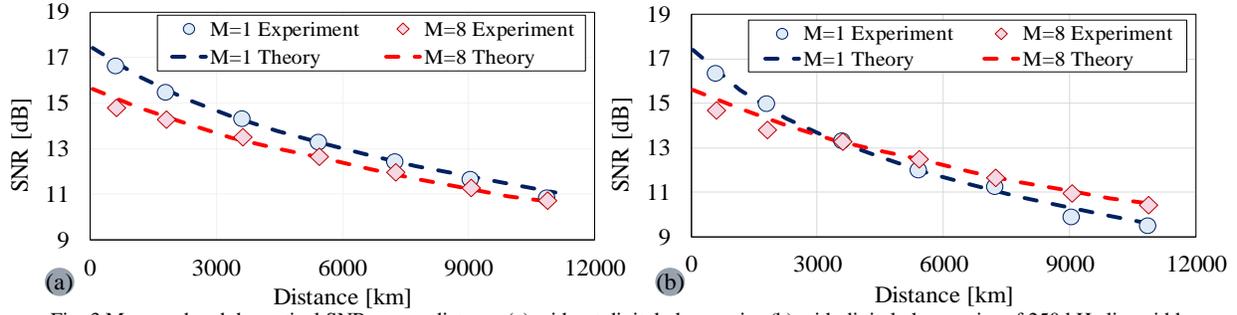
Fig. 3 Measured and theoretical SNR versus distance (a) without digital phase noise (b) with digital phase noise of 250 kHz linewidth

$$SNR = \left[\frac{1}{\eta s_0} + \frac{\pi c DRL\Delta\vartheta}{2\zeta M f_0^2} + \frac{3}{2^{2/3}}\left(\frac{a_{nli}N_0^2}{M^2}\right)^{1/3}\right]^{-1}$$

With $\eta$ is performance improvement of SNR ceiling of MC w.r.t SC, $s_0$ is the reference B2B single-carrier SNR ceiling, $f_0$ is the channel frequency, $c$ is the speed of light, $D$ is the fiber dispersion coefficient, $R$ is the single-carrier baudrate, $L$ the transmission distance, $\Delta\vartheta$ is the laser linewidth, $a_{nli}$ the nonlinear coefficient, $N_0$ the ASE noise variance assuming 5 dB of noise figure, and $M$ is the number of digital subcarriers. Refs. [7] and [12] reported that the analytical model for the EEPN [6] overestimates its effect. According to [12] we introduced the coefficient $\zeta\sim3$ to account for this overestimation. The $a_{nli}$ values come from Fig. 1 b) and the $\Delta\vartheta$ was found to be 60 kHz by fitting. After 605 km, we observe that the SC format outperforms the MC format where the measured SNR for SC is 1.7 dB higher compared to MC. This difference is attributed to the higher sensitivity of MC to transceiver imperfections [9]. The evolution of SNR shows equal distance-dependent noise contribution for MC and SC. The performance of these formats converges with distance, differing by less than 0.25 dB after 10890 km demonstrating a same tolerance to nonlinearities and EEPN and confirming the expectations from simulations considering a 60 kHz laser linewidth. Moreover, to assess the impact of the local oscillator (LO) laser linewidth, we digitally emulated a larger LO linewidth by adding phase rotations on the real-time sampled waveforms of the coherent receiver. The phase noise process is assumed to be a Wiener process of 250 kHz linewidth, larger than the 60 kHz linewidth of our LO. Results are illustrated in Fig.3. b) depicting the SNR as a function of the distance for this digitally enhanced LO linewidth. We observe a better resilience of MC signal to the additional phase noise. At the longest distance of 10890 km, the SNR of SC degrades by 1.3 dB for such larger LO linewidth while the MC worsens by only 0.25 dB. It is worth noting that the performance with larger linewidth remains worse than with narrower linewidth independently from using MC or SC. These results indicate that commercial standard ITLA LO sources exhibit sufficiently good linewidths to provide optimal performance of SC and MC over an ultra-long accumulation of chromatic dispersion.

## 4. Conclusions

In this paper we assessed the impact of nonlinearities on both single-carrier (SC) and digital multi-carrier (MC) signals at 120 GBd with probabilistically shaped formats. Numerical investigation as well as transmission results showed similar behavior of MC and SC in the presence of nonlinear effects at long distance. Moreover, though MC signals can mitigate the penalties arising from large linewidth lasers over long-haul distances with large cumulated dispersion, standard ITLA sources provide low enough linewidth to reach optimal performance for both MC and SC signals.

*Acknowledgments: The work of Stenio M. Ranzini is supported by the European Union H2020 under MSCA GA 766115.*

## 5. References

[1] F. Buchali et al., "Experimental demonstration of capacity increase and rate-adaptation by probabilistically", ECOC 2015, Paper PDP.3.4
[2] A. Ghazisaeidi et al., "Advanced C+ L-band transoceanic transmission systems based on PS PDM-64QAM", JLT vol. 35, no. 7, 2017.
[3] H. Sun et al. "800G DSP ASIC Design Using Probabilistic Shaping and Digital Sub-Carrier Multiplexing", in JLT, vol. 38, no. 17, 2020.
[4] A. Arnoult, et al, "Field Trial Demonstration over Live Traffic Network of 400 Gb/s Ultra-Long Haul and 600 Gb/s Regional Transmission", ECOC 2020, paper TU2E.4.
[5] P. Poggiolini et al., "Analytical and Experimental Results on System Maximum Reach Increase Through Symbol Rate Optimization", in JLT, vol. 34, no. 8, 2016.
[6] W. Shieh et al, "Equalization-enhanced phase noise for coherent detection systems using electronic digital signal processing", in Opt. Express, vol. 16, no. 20, 2008.
[7] A. Arnoult et al. "Equalization Enhanced Phase Noise in Coherent Receivers: DSP-Aware Analysis and Shaped Constellations" in JLT vol 37 no. 20, 2019
[8] A. Carbo, et al., "Experimental Analysis of Non Linear Tolerance Dependency of Multicarrier Modulations versus Number of WDM Channels", OFC 2016 paper Tu3A.6.
[9] F. Buchali et al. "Study of electrical subband multiplexing at 54 GHz modulation bandwidth for 16QAM and PS64QAM", ECOC 2016.
[10] F. Buchali, et al. "128 GSa/s SiGe DAC implementation enabling 1.52 Tb/s single carrier transmission." OFC 2020, paper Th4C.2.
[11] G. Bosco, et al. "Impact of the transmitter IQ-skew in multi-subcarrier coherent optical systems." OFC 2016, paper W4A.5.
[12] A. Carbo et al. "Experimental Characterization of EEPN in Transoceanic Transmission Systems", ECOC 2019. paper WE1A.2.